\documentclass[aps,prd,twocolumn,superscriptaddress,groupedaddress,preprintnumbers]{revtex4}

\usepackage[english]{babel}
\usepackage{array}
\usepackage{tabulary}
\usepackage{amsfonts}
\usepackage{amsmath}
\usepackage{cancel}
\usepackage{mathcomp}
\usepackage{graphicx}
\usepackage{color}

\def\beq{\begin{equation}}
\def\eeq{\end{equation}}
\def\baq{\begin{eqnarray}}
\def\eaq{\end{eqnarray}}

\begin{document}

\title{Coupled dark energy and dark matter from dilatation anomaly}

\preprint{HD-THEP-10-22 $\quad$ NORDITA-2010-106}

\author{J.~Beyer$^a$, S.~Nurmi$^{a,b}$ and C.~Wetterich$^a$ \\
{\it $^a$Institut f\"ur Theoretische Physik,   
Universit\"at Heidelberg,   
Philosophenweg 16, 69120 Heidelberg, Germany \\
$^b$NORDITA, Roslagstullsbacken 23, SE-106 91 Stockholm, Sweden }}

\begin{abstract}
Cosmological runaway solutions may exhibit an exact dilatation
symmetry in the asymptotic limit of infinite time. In this limit, the
massless dilaton or cosmon could be accompanied by another massless
scalar field - the bolon. At finite time, small time-dependent masses
for both the cosmon and bolon are still present due to imperfect
dilatation symmetry. For a sufficiently large
mass the bolon will start oscillating and
play the role of dark matter, while the cosmon is responsible for dark
energy. The common origin of the mass of both fields leads to an effective
interaction between dark matter and dark energy. Realistic cosmologies
are possible for a simple form of the effective
cosmon-bolon-potential. We find an inverse bolon mass of a size where it
could reduce subgalactic structure formation.
\end{abstract}

\maketitle


\section{Introduction}
Dilatation symmetry and its anomaly could play an important role for cosmology \cite{Wetterich:1987fm}. Models with a dilatation symmetric fixed point
could provide a dynamical solution for the cosmological constant
problem \cite{Wetterich:1994bg, Wetterich:1987fm,
  Wetterich:2002wm}. As one of the most characteristic features such models have predicted the presence of a homogeneous dark energy component \cite{Wetterich:1987fm}, long before its observational discovery. Recent investigations of higher dimensional settings have shed new light on such theories \cite{Wetterich:2008bf}. It has been shown that dimensional reduction of a dilatation symmetric quantum effective
action in higher dimensions leads to four-dimensional models with a vanishing cosmological
constant. If the cosmological solution approaches the
dilatation symmetric fixed point in the limit $t\rightarrow \infty$,
such models can naturally give rise to an asymptotically vanishing
cosmological constant. In such a scenario the observed particle masses are due to the
spontaneous breaking of dilatation symmetry by the cosmological or
vacuum solution. At the fixed point one will therefore encounter a
massless goldstone boson - the dilaton. For finite time, the dilatation symmetry is
broken by anomalous terms which generate a small time-dependent mass
for the dilaton. This picture can give
rise to a quintessence cosmology where the dilaton rolling towards
the fixed point plays the role of the scalar "cosmon"-field with a slowly
decreasing mass \cite{Wetterich:1987fm, Wetterich:1994bg}. This higher dimensional dilatation symmetric setup
naturally leads to scaling solutions where the cosmon mass
tracks the Hubble parameter.

First investigations of the manifold of extrema of a dilatation
symmetric higher-dimensional quantum effective action often reveal the
presence of additional massless scalar fields besides the dilaton
\cite{Wetterich:2008bf}. These may correspond to a change of the
characteristic length scale of internal space - the radion -, or other
changes in geometry, similar to the moduli fields in string theory. In
this work we will concentrate on one such field and name it the
"bolon'', since it will be ultimately responsible for dark matter and
therefore for the emergence of structure (i.e. "lumps" or "bola") in the cosmos.

We start with a quick revision of the key concept of dilatation symmetry. Then we will investigate the coupled system of cosmon and
bolon and show that it can reproduce the standard cosmological evolution at the background level for rather simple potentials. After a radiation dominated period, during which the cosmon and bolon act as early dark energy and play a subdominant role, the bolon starts to oscillate around a partial potential minimum. Its oscillation energy (potential and kinetic) is diluted as non-relativistic matter and thus it will eventually become dominant, enforcing a transition to a matter dominated period. After the transition the fluid equations for the energy density of bolon fluctuations obey the standard form for cold dark matter, with a small coupling to the cosmon. Thus the simple model with two scalars describes a cosmology with coupled dark matter and dark energy.

We further proceed to slight modifications of the simplest potential for which dark energy finally dominates the energy density of the universe. This can be achieved by an effective stop in the cosmon evolution, induced either by a characteristic change in the scalar potential or a leaping kinetic term, or if the scaling behaviour of the cosmon gets terminated by a cosmic trigger event. An example of the last scenario is given by the growing neutrino quintessence model \cite{Amendola:2007yx}.

\section{Dilatation symmetry}
Dilatations correspond to a rescaling of each dynamical field of a given theory corresponding to an appropriate scaling dimension. For a theory involving a metric ${\rm g}_{\mu \nu}$ and a scalar field $\xi$ in $d$ dimensions this means
\beq
{\rm g}_{\mu \nu} \rightarrow \alpha^2 {\rm g}_{\mu \nu} \, , \quad \xi \rightarrow \alpha^{-((d-2)/2)} \xi \, .
\eeq
Any dilatation symmetric quantum effective action $\Gamma$ is then given by an action
\beq
\Gamma = \int d^4x \, {\rm g}^{1/2} \mathcal{L} \, 
\eeq
with 
\beq
\mathcal{L} \rightarrow \alpha^{-d} \mathcal{L} \, 
\eeq
under dilatations. This puts severe restrictions on the allowed interactions in $\mathcal{L}$. Dilatation symmetry ensures the absence of any explicit mass scales of the model. These restrictions have particularly interesting consequences in a higher dimensional setting. For $d > 6$ no polynomial potential for the scalar field is allowed, and dimensional reduction of such a higher dimensional dilatation symmetric theory gives rise to a rather generic class of effective four-dimensional theories with a vanishing cosmological constant \cite{Wetterich:2008bf}. This finding has been generalized to arbitrary dilatation symmetric $\Gamma$, for example based only on the metric without a scalar field \cite{Wetterich:2008bf}.

In our scenario we consider theories with a dilatation symmetric fixed point, which corresponds to field configurations reached in the limit of cosmic time going to infinity. One possibility is to start with a generic action $\Gamma$, not necessarily dilatation symmetric, and to consider a family $\Gamma_\kappa$ obtained by a rescaling
\beq
\Gamma_\kappa [{\rm g}_{\mu \nu}, \xi] = \Gamma[\kappa^{-2} {\rm g}_{\mu \nu}, \kappa^{d(d-2)/2} \xi] \, .
\eeq
This becomes dilatation symmetric in the limit $\kappa \rightarrow \infty$. One may imagine that a dynamical runaway solution drives $\kappa$ effectively to infinity for $t \rightarrow \infty$, for example by a monotonic increase of the cosmological value of the scalar $\xi$.

Dimensional reduction of such an action leads to infinitely many fields in four dimensions. We consider only massless or light fields which play a role in cosmology after a possible inflationary period. They are described by an effective theory, encoded in the four-dimensional, dimensionally reduced action, plus the metric and the particles of the standard model of particle physics. The latter will dominate the radiation dominated era, while in the matter dominated era the small component of baryons is subleading and can be neglected for the overall picture of dark matter. Our starting point for the effective action of two scalars $\varphi$ and $\chi$ is a standard kinetic term and a potential $V(\varphi, \chi)$. Dilatation symmetry at the fixed point enforces the common potential of the two fields to vanish for $\varphi \rightarrow \infty$ \cite{Wetterich:2008bf}.

\section{The model}
Away from the dilatation symmetric fixed point the
four-dimensional quantum effective action will contain an effective potential
$V(\varphi, \chi)$ for the cosmon field $\varphi$ and the bolon field
$\chi$, generated by dilatation anomalies. The formulation in terms of the quantum effective action means that no further quantum corrections to the potential are present. If we were to start at the classical theory, various quantum corrections would appear in the process of quantization. They may in fact be responsible for the existence of a fixed point. In this case the effective dilatation symmetry at the fixed point is due to quantum fluctuations (rather than being destroyed by them) and may be realized even for a classical action without dilatation symmetry. In the vicinity of a fixed point asymptotic dilatation symmetry guarantees that the potential vanishes as the fixed point is approached for $\varphi \rightarrow \infty$. It also ensures a vanishing mass at the fixed point. The bolon mass also vanishes at the fixed point.

We work within the effective four-dimensional theory (obtained
after dimensional reduction) and use the Einstein frame (with fixed Planck mass). Furthermore we
assume first a normalization of $\varphi$ and $\chi$ such that both
have standard kinetic terms. In this frame the $\varphi$-dependence of
the potential is typically of exponential shape
\cite{Wetterich:1987fm} and we start with a simple model where
  \beq
  \label{V}
  V=M^4\left[\left(\frac{\mu}{M}\right)^{A} e^{-\alpha\varphi/M}
  +\left(\frac{\mu}{M}\right)^{B}
  e^{-2 \beta\varphi/M} \left( \frac{\chi}{M} \right)^2\right]\ .
  \eeq
Here $M$ is the four-dimensional
effective Planck mass and $\mu\ll M$ is the scale of anomalous
dilatation symmetry breaking. We assume the dimensionless
constants $A$, $B$, $\alpha$ and $\beta$ to be all positive. The
"anomalous dimensions'' $A$, $B$ characterize the impact of explicit
mass scales of the model (violation of dilatation symmetry) on the
four-dimensional world. The asymptotic solution will correspond to
$\varphi \rightarrow \infty$ where the potential $V$ vanishes and
dilatation symmetry becomes exact. Then also
the mass matrix for the cosmon and bolon, given
by the second derivatives of $V$, approaches zero. The simple quadratic form for $\chi$
is assumed to be a good approximation for $\chi$ close to a partial
minimum of the potential in this direction. Small deviations of $\chi$
from the partial extremum will be sufficient for the understanding of
late cosmology. For very early cosmology the quadratic approximation
for $\chi$ may become insufficient and we discuss generalizations
below. There we also briefly address the possibility that the
$\varphi$-dependence of the partial minimum for $\chi$ cannot be
neglected (as for the potential (\ref{V})).

An important ingredient for our setting concerns the observation that
the approach to the fixed point at $\varphi \rightarrow \infty$, $\chi
\rightarrow 0$ can be characterized by a different scaling for the
cosmon and the bolon directions. For $\chi$ at its minimum the size of
the "anomalous potential'' is dictated by the parameters $A$ and
$\alpha$. Different parameters $B$ and $\beta$ may describe the
scaling of $V$ away from the partial minimum, e.g. for $\chi \neq
0$. The characteristic scaling behaviour and its motivation by the
properties of solutions in the vicinity of a higher dimensional dilatation
symmetric fixed point distinguishes our approach from other
settings where dark energy and dark matter are described by two scalar
fields, as realized, for example, in a model for the axion and the
cosmon or in the work of ref. \cite{Matos:2000ss, Sahni:1999qe} and
\cite{Anchordoqui:2007sb, Axenides:2004kb, Henriques:2009hq,
  Mainini:2006pf}. Due to the factor $\text{exp}\left( -2 \beta
  \varphi / M \right)$ the mass of the bolon field will depend on the
value of the cosmon field. If the bolon plays the role of dark matter
we expect a coupling between dark energy and dark matter $\sim \beta$,
as discussed previously in models of coupled quintessence \cite{Wetterich:1994bg, Amendola:1999er}. We
emphasize that a cosmon-bolon coupling is necessary in our setting
since the bolon mass is assumed to vanish for the asymptotic solution.

\section{Cosmic evolution}
To simplify the analysis of the dynamics, we perform a shift
$\varphi\rightarrow \varphi-(MA/\alpha) \, {\rm
    ln}\,(\mu/M)$ which corresponds to $A \rightarrow 0$ and $B
  \rightarrow \widetilde{B}=B-\beta\,A/\alpha$ in
eq. (\ref{V}). This rescaling shows that our model depends effectively on three parameters, namely the dimensionless couplings $\alpha$ and $\beta$ and the effective scale for the bolon mass, which involves the parameter $(\mu / M)^{\tilde{B}}$.
We define the
energy densities
  \baq
  \label{rho_phi}
  \rho_{\varphi}&=&\frac{1}{2}\dot{\varphi}^2+M^4 e^{-\alpha\varphi/M}\ ,\\
  \label{rho_chi}
  \rho_{\chi}&=&\frac{1}{2}\dot{\chi}^2+M^4\left(\frac{\mu}{M}\right)^{\widetilde{B}}
  e^{-2 \beta\varphi/M} \left( \frac{\chi}{M} \right)^2\
  ,
  \eaq
which will be associated with dark energy and dark matter components
respectively. In addition to $\rho_{\chi}$ and
$\rho_{\varphi}$, we assume that the universe also contains a homogeneous
radiation component $\rho_r$ which dominates at early times. 

The $\varphi$-dependent mass of the bolon, reflected by the factor ${\rm exp}(-2 \beta \varphi / M )$ in eq. (\ref{rho_chi}), will lead to a coupling between dark matter and dark energy of strength $\beta$, similar to generic models models of coupled quintessence \cite{Wetterich:1994bg, Amendola:1999er}. We would like to point out that in our model the coupling between the bolon and the cosmon does not arise from matter particles coupling to a metric related to the Einstein frame metric by a conformal transformation as e.g. in generalized Jordan-Brans-Dicke theories \cite{Damour:1990tw}. It rather emerges as a feature of the common cosmon-bolon potential from the breaking of dilatation symmetry and is crucial to restore dilatation symmetry at the fixed point. Possible couplings of the cosmon to other forms of matter as baryons or neutrinos are not directly related to the coupling $\beta$ appearing in eq. (\ref{V}). In particular, there is no reason to expect a baryon-cosmon coupling of equal strength as the coupling between dark matter and dark energy. Limits on the time variation of couplings and violations of the weak equivalence principle imply for most models that the baryon-cosmon coupling must be much smaller than the value of $\beta$ discussed in this paper \cite{Wetterich:1987fm, Wetterich:1994bg, Wetterich:2002ic}. Nevertheless it is possible to add a baryonic fluid which can be either uncoupled, very weakly coupled or coupled with gravitational strength to the dilaton without qualitatively changing our conclusions. In the last case, however, one would need to invoke some kind of mechanism to avoid constraints from local tests of gravity, like the Damour-Polyakov effect \cite{Damour:1994zq, Brax:2010gi} or the chameleon mechanism \cite{Khoury:2003aq, Hinterbichler:2010wu}.

For $\alpha\neq 2 \beta$, the two terms in (\ref{V}) decay at
different rates as $\varphi\rightarrow\infty$. The qualitative
features of the dynamics in the early radiation dominated epoch
depend on the ratio of $\alpha$ and $\beta$ and the magnitude of the
prefactor $(\mu/M)^{\widetilde{B}}$. We start by discussing
solutions for which the potential (\ref{V}) is dominated by
the first term in the early radiation dominated universe and the
second term becomes dominant at late times causing a transition into
a matter dominated epoch. The growth of the second term in
(\ref{V}) relative to the first one requires $2 \beta <
\alpha$, but we will actually impose a stronger condition, $\beta\ll
\alpha$. This will also ensure that the resulting dark matter-dark
energy interaction remains compatible with cosmological observations. The dominance of the
first term in (\ref{V}) (small positive values of $\varphi$) requires
$({\mu}/{M})^{\widetilde{B}}\chi^2\ll M^2$.

In the early radiation dominated universe, we then have $V\simeq M^4
e^{-\alpha\varphi/M}$ and the system approaches rapidly the well known attractor
solution \cite{Wetterich:1987fm, Wetterich:1994bg}
  \beq
  \label{early_scaling}
  \varphi=-\frac{M}{\alpha}\,{\rm
  ln}\left(\frac{4H^2}{M^2 \alpha^2}\right)\ ,\qquad\rho_{\varphi}=\frac{12}{\alpha^2}\,H^2M^2\ .
  \eeq
The behaviour of $\chi$ depends on the magnitude of its effective
mass which obeys for the scaling solution (\ref{early_scaling})
  \beq
\label{DilatonMass}
  m_{\chi}^2=2M^2\left(\frac{\mu}{M}\right)^{\widetilde{B}}\Omega_{\varphi}^{2
    \beta/\alpha}
  \left(\frac{H}{M}\right)^{4\beta/\alpha}\ ,
  \eeq
where $\Omega_{\varphi}=\rho_{\varphi}/(3M^2H^2)=4/\alpha^2$. For $2
\beta < \alpha$, the mass grows relative to the Hubble rate,
$m_{\chi} /H\propto t^{1-2 \beta/\alpha}$. For early cosmology one has
$H \gg m_\chi$ and the
field $\chi$, starting from some initial configuration $\chi_{\rm
  in}$, $\dot{\chi}_{\rm in}$, settles rapidly to a constant value $\chi_0$.
Subsequently, it remains nearly frozen at $\chi_0$ until $m_\chi \sim H$, when $\chi$
starts to oscillate.

Assuming $\rho_{\chi}$ to be still subdominant compared to
$\rho_{\varphi}$ at this point, we can use the scaling solution
(\ref{early_scaling}) to obtain $\dot{m}_{\chi}/m_{\chi}^2 =
-4(\beta/\alpha)H/m_{\chi}$. For $\beta\ll \alpha$, $m_{\chi}$ soon
becomes nearly constant in a time scale of one oscillation cycle
$1/m_{\chi}$. As a consequence, the bolon $\chi$ oscillates in an effectively
quadratic potential with a slowly decreasing mass term. Its energy
density behaves almost as that of a non-relativistic matter
component \cite{Turner:1983he} and eventually comes to dominate over $\rho_{\varphi}$ and
over the radiation component.

For an estimate of the onset of oscillations we define $t_{\rm osc}$
 by the condition $H(t_{\rm osc}) = m_{\chi} (t_{\rm osc})$. In the absence of a coupling between $\chi$
and $\varphi$, $\beta \rightarrow 0$, we obtain $H(t_{\rm osc}) =
\sqrt{2}M (\mu/M)^{\tilde{B}/2}$. The time of matter radiation equality is
approximately given by
  \beq
  \label{H_eq}
  \frac{H_{\rm eq}}{M} \sim
  \left(\frac{\mu}{M}\right)^{\widetilde{B}/2}\left( \frac{\chi_{\rm
        eq}}{M} \right) \, .
 \eeq
 Here $\chi_{\rm eq}$ denotes the bolon-amplitude at
$t_{\rm eq}$ and obeys
\beq
\frac{\chi_{\rm eq}}{M} \sim \left( \frac{\chi_0}{M} \right)^{4} \, 
  \eeq
if $\chi_0 \ll M$. For a weakly coupled system with $\beta/\alpha\ll 1$, the energy
density stored in the oscillations $\rho_{\chi}\sim
(1/2)m_{\chi}^2\chi^2$ dilutes slightly faster than $a^{-3}$ due to
the decrease of the mass $m^2_{\chi}(\varphi)$. Our
numerical analysis reveals that eq. (\ref{H_eq}) gives a fairly accurate estimate
in the weakly coupled case as well. Realistic cosmologies require  matter-radiation equality to happen
at $H_{\rm eq}/M \approx 8 \times 10^{-56}$. For initial conditions
leading to $\chi_0$ of the order $0.1 M$ this requires a small value of
$(\mu / M)^{\tilde{B}/2}$, say $\mu \approx 10^{8}$ GeV for
$\tilde{B} = 10$. (Still the value of $\mu$ is much larger than all the
energy scales of the standard model of particle physics - smaller
$\mu$ can be obtained for smaller $\tilde{B}$.)

After the onset of the bolon oscillations, the evolution equations for
the system can be expressed in the form
  \baq
  \label{EC1}
  \ddot{\varphi}+3H\dot{\varphi}-\alpha M^3
  e^{-\alpha\varphi/M}&=&\frac{\beta}{M}\,\rho_{\chi} \, ,\\
\label{EC2}
  \dot{\rho}_{\chi}+3H\rho_{\chi}&=&-\frac{\beta}{M}\,\rho_{\chi}\dot{\varphi}
  \, , \\
  \dot{\rho}_{\gamma}+4H\rho_{\gamma}&=&0 \, ,\\
  \label{EC4}
  3M^2H^2&=&(\rho_r+\rho_{\chi}+\rho_{\varphi})\, ,
  \eaq
where we have averaged over an oscillation cycle and used the
results $\langle\dot{\chi}^2\rangle=2 \langle
M^2(\mu/M)^{\widetilde{B}}e^{-2
\beta\varphi/M}\chi^2\rangle=\rho_{\chi}$, valid for
$H\ll m_{\chi}$ and $(\dot{m}_{\chi} /m_{\chi}) \ll H$. 
For these "late" times the cosmon-bolon system bears much resemblence to other theories where a quintessence field couples to dark matter through the trace of the energy-momentum tensor, e.g. chameleon models \cite{Khoury:2003aq} or the model investigated in \cite{Comelli:2003cv}. (The main difference between the cosmic evolution described in \cite{Comelli:2003cv} and our scenario - besides the obvious differences for early cosmology - is that the authors of \cite{Comelli:2003cv} describe a "freezing" solution. There the universal attractor is characterized by an accelerated expansion and the cosmon field remains frozen until very recently and then starts to dominate the cosmic evolution. While this type of solution is also present in our scenario for a suitable choice of parameters and initial values, it is very sensitive to the choice of initial conditions. For this reason we will focus here on scaling solutions for the cosmon field during both the matter and the radiation dominated eras. We will see in section \ref{sec5} how this can ensure insensitivity to initial conditions. There we also discuss possibilities how to break this behaviour and reach cosmic acceleration.)

The dynamics of the system described by eq. (\ref{EC1}) to eq. (\ref{EC4}) was analyzed in \cite{Wetterich:1994bg,
  Amendola:1999er}. For
$\beta/\alpha\ll 1$, it admits an effectively matter dominated
attractor solution with
  \beq
  a\propto t^{1-\beta/\alpha}\ ,\qquad \rho_{\varphi}=3 M^2 H^2
  f(\alpha,\beta) \ ,
  \eeq
where $f=(18+6\beta^2-6\beta\alpha)/(6(\alpha-\beta)^2)$. For
$\alpha \gg 1$, as required by $\rho_{\varphi}\ll \rho_{r}$ in
the early scaling epoch, stability and existence os the solution is guaranteed for $4\beta < \alpha < \beta+3/\beta$. Expanding
$f(\alpha,\beta)$ around $\beta/\alpha=0$, we find
  \beq
  \label{CoupledScalingSolution}
  \Omega_{\varphi}=\frac{3}{\alpha^2}-\frac{\beta}{\alpha}\left(1-\frac{6}{\alpha^2} \right)
    + {\cal
  O}(\beta^2/\alpha^{2})\, ,
  \eeq
which shows that $\Omega_{\varphi}$ decreases slightly as compared to
the early radiation dominated epoch. For $\alpha\sim {\cal O}(10)$,
$\rho_{\varphi}$ constitutes a few percent of the total energy
density in the matter dominated epoch. 
\begin{figure}[h]
  \centering
  \includegraphics[width=0.98 \columnwidth]{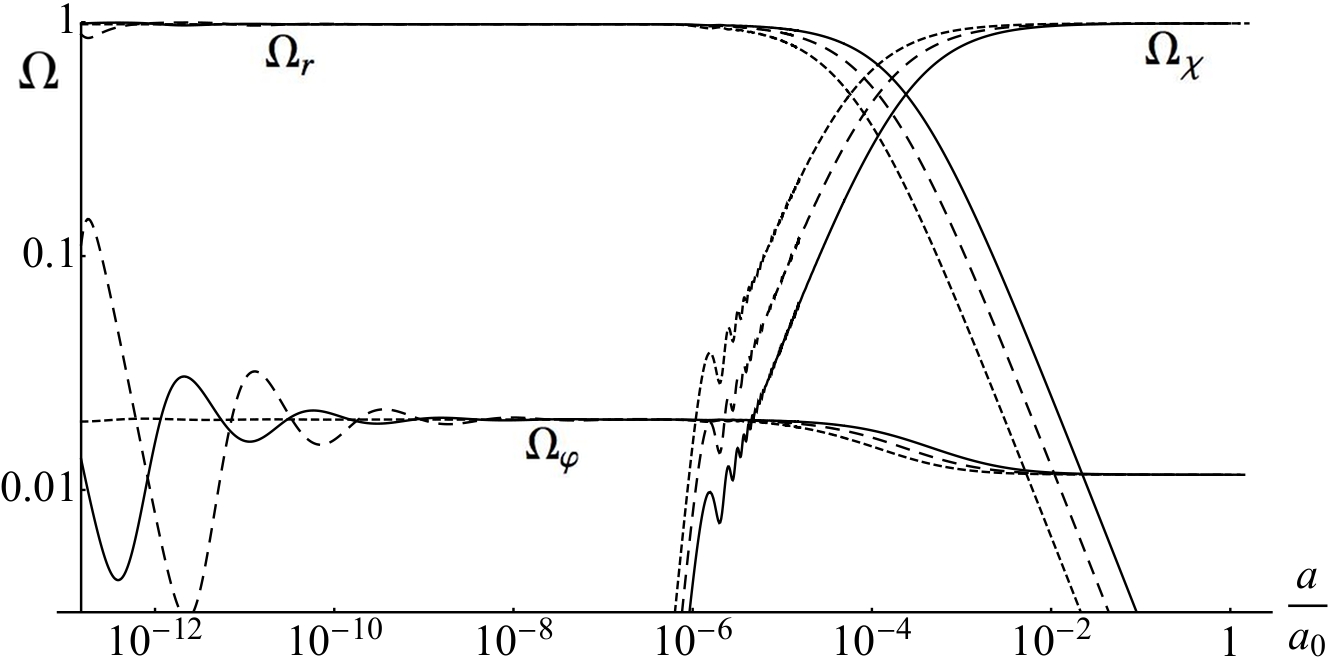}
  \caption[Density parameters]{Evolution of the density
    parameters
$\Omega_r$,
$\Omega_{\chi}$ and
$\Omega_{\varphi}$ for $\alpha=\sqrt{2} \times 10$, $\beta = 0.05$,
$\left( \frac{\mu}{M} \right)^{\tilde{B}} = 10^{-100}$ and
$\dot{\chi}_{in} = 0$. We show three different values for
$\chi_{in}/M$: 0.1
(solid), 
 $\sqrt{2} \times 0.1$
(dashed) 
 and 0.2. We also vary $\Omega_{\varphi}$ ($z=10^{13}$).}
  \label{RMEquality}
\end{figure}

At this stage the coupled cosmon-bolon system describes a cosmology with dark
matter, radiation and a small constant fraction of early dark
energy. A realistic cosmology requires a dark matter - dark energy
crossover where after redshift $z \approx 5$ dark energy increases
from a few percent to its
present fraction $\Omega_\varphi > 0.7$. This can be achieved by an
effective stop or strong slowing down of the evolution of $\varphi$. Such a stop may be induced by a cosmological
trigger event - as neutrinos becoming non-relativistic
\cite{Amendola:2007yx} or scalar backreaction due to structure
formation \cite{Wetterich:2003kr}, or by a qualitative change of the cosmon
potential or the kinetic term for $\varphi$ exceeding a characteristic
value \cite{Hebecker:2000au}. 

\section{Modifications of the scalar potential}
\label{sec5}

\subsection{bolon scaling in the early universe}
For the potential (\ref{V}) quadratic in $\chi$,  the time of
matter-radiation equality $t_{\rm eq}$ depends on both the model
parameter $(\mu/M)^{\tilde{B}}$ and the initial conditions for $\chi$,
see eq.~(\ref{H_eq}).  This is illustrated in
fig.~\ref{RMEquality}. The memory of the initial conditions for
$\varphi$ disappears as the cosmon hits the attractor solution.  For a
very wide range of initial conditions this happens long before the
transition to matter domination, as demonstrated in
fig.~\ref{RMEquality}.  A dependence of $t_{\rm eq}$ (and the final
value of $\rho_{\chi}a^3$) on the initial conditions for $\chi$ is
common to other dark matter models like the axion. While realistic
cosmology can be obtained for rather natural looking initial
conditions, e.g. for $\chi_0\sim M$ with $(\mu/M)^{\tilde{B}}\sim
10^{-100}$, it is noteworthy that the initial energy density
$\rho_{\chi}$ needs to be considerably smaller than $\rho_{\varphi}$ (and
$\rho_{\rm r}$) for the potential (\ref{V}) to work. One could presume that the fields  $\varphi$ and $\chi$
have been excited at some point after inflation in a similar
process, possibly connected to the production of baryonic
matter and radiation. In this case initial energy densities
of the same order of magnitude would seem more natural.
Furthermore, assuming the quadratic dependence on $\chi$ in eq.~(\ref{V}) to remain valid all the way to the inflationary epoch, one would run into
problems with the excessive isocurvature perturbations
arising from the fluctuations of the light bolon.

We therefore consider eq. (\ref{V}) to be a valid approximation only for late
cosmology (i.e. field configurations close to the fixed point). Indeed, there is no reason to assume a quadratic dependence on
$\chi$ for large values, as relevant for early cosmology. In this regime the potential may be much steeper, typically involving $\chi$
exponentially. Steep potentials typically admit tracker
solutions for which $\rho_\chi$ will either scale like the dominant
background fluid (for an exponential $\chi$-dependence) or
slowly catch up to it. This could lead to a setting where the time of
matter-radiation equality only depends on model parameters and not any
longer on the initial conditions for $\chi$. In a first period $\chi$
will then rapidly evolve until it settles at a value $\bar{\chi}_0$. There is sits until $\rho_\chi$ has caught
up with the energy density of the tracker-solution, which it will
follow from this point on. Once $\chi$ has reached values sufficiently close
to its minimum the quadratic approximation for the $\chi$-dependence may
be used. This behaviour occurs for a large range of initial conditions
and guarantees that the bolon enters the flat part of the potential
with predetermined values of $\chi$ and $\dot{\chi}$, which now depend only
on model parameters.

A simple example for such a mechanism can be obtained by replacing $\chi^2 / M^2$ in
eq. (\ref{V}) by $({\rm cosh}(\lambda\chi/M)-1)$, a form
considered in the context of uncoupled scalar dark matter in ref.
\cite{Matos:2000ss,Sahni:1999qe}. This potential has an
asymptotically exponential shape for large $\chi$-values, allowing
for an attractive scaling solution in the uncoupled case
$\beta = 0$. Our numerics reveal that the qualitative behaviour is maintained in the weakly
coupled case with small $\beta$. The range of initial conditions
resulting in realistic cosmologies for this modification is rather
large, including equipartition of the energy-densities $\Omega_\chi
(t_{\rm in}) \approx \Omega_r (t_{\rm in})$. Fig. \ref{ScalingBreaking}
demonstrates that this wide range of initial conditions only
influences early cosmology. Close to matter radiation equality
essentially no memory of the initial conditions survives. For late
cosmology the behaviour is very similar to the potential (\ref{V}), but
now with an effective value $\chi_0$ that is determined by model
parameters and no longer by initial conditions. We would like to point out that the price we pay for the independence of initial conditions is, as usual, the introduction of an additional parameter (in this simple model $\lambda$). It determines at which field value the bolon potential makes a transition from a "steep" to a "flat" regime. The parameter needs to be adjusted to give a realistic cosmic evolution with the correct "timing" of matter-radiation equality.

\subsection{Breaking the scaling behaviour with a cosmon-dependent minimum}
\begin{figure}[h]
\includegraphics[width = 0.98 \columnwidth]{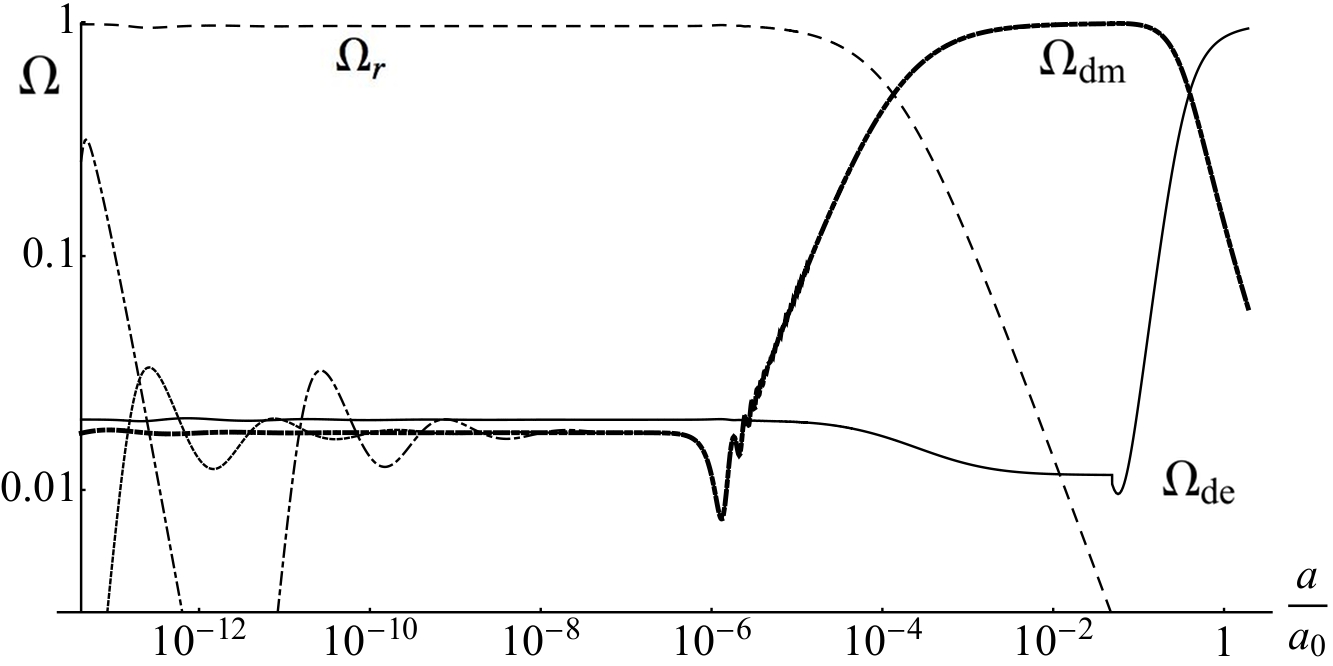}
  \caption[Modification 1]{Dark matter - dark energy crossover for parameters $\gamma = 0$, $\alpha=\sqrt{2} \times 10$,
$\beta = 0.05$,
$\left( \frac{\mu}{M} \right)^{\tilde{B}} = 10^{-96}$, $c=10$ and
$\varphi_0 = 8.7 M$. The potential for large $\chi$ values was modified
to have exponential shape by replacing $\chi^2$ by $({\rm
  cosh}(\lambda \chi) - 1)$, $\lambda = 15$. The bold curve for
$\Omega_{\rm dm}$ reflects a large range of initial conditions after
inflation. We demonstrate the insensitivity to initial values by
setting new ``initial conditions'' at $z=13$ with $\chi$ between $4M$
and $5M$, $\dot{\chi} = 0$ (dashed curves).
}
\label{ScalingBreaking}
\end{figure}

A further interesting generalization of our model potentially concerns
a $\varphi$-dependence of the partial minimum for $\chi$ according to
substituting $\chi^2$ in eq.~(\ref{V}) by $e^{-2\gamma
    \chi/M}\,\left(\chi - g (\varphi) \right)^2$.
This could lead to a late time accelerated expansion if $g(\varphi)$ has a characteristic shape where it changes its form
abruptly once $\varphi$ passes a characteristic value $\varphi_0$.
We demonstrate this in fig.~\ref{ScalingBreaking} for a simple toy
model where $g(\varphi) = c (\varphi - \varphi_0)
\Theta(\varphi-\varphi_0)$, $\Theta$ being the Heaviside function. One observes a crossover from matter
domination to dark energy domination in the present cosmological
epoch.

If the partial minimum for the bolon depends on $\varphi$ the
separation of the combined bolon cosmon-bolon energy density into dark
matter and dark energy is less straightforward. A reasonable
approximation for $\varphi>\varphi_0$  is given by
\begin{align}
  \rho_{dm} &= \frac{1}{1+c^2} (\dot{\chi} - c \dot{\varphi} )^2 +
  V(\varphi,\chi) - V(\varphi,g(\varphi)) \, , \nonumber \\
\rho_{de} &= \frac{1}{1+c^2} (c \dot{\chi} + \dot{\varphi} )^2 +
V(\varphi,g(\varphi)) \, .
\end{align}
The time averaged evolution equations for these quantities are given
by eq.~(\ref{EC1}) and (\ref{EC2}), but with $\rho_\chi$, $\rho_\varphi$,
$\alpha$, $\beta$ and $\varphi$ replaced by $\rho_{dm}$, $\rho_{de}$,
$\alpha_c = \alpha/\sqrt{1+c^2}$, $\beta_c =
(\beta + c \gamma)/\sqrt{1+c^2}$ and the approximation for the
cosmon, $\phi = (c
\chi + \varphi)/\sqrt{1+c^2}$, respectively.
For
$\varphi>\varphi_0$ the scaling solution
(\ref{CoupledScalingSolution}) remains no longer valid - the
asymptotic behaviour is now governed by $\alpha_c$ and the new
coupling $\beta_c$.
For an appropiate choice of parameters the only stable fixed point of
this system is an
accelerated expansion \cite{Amendola:1999er}. We show an example in
fig.~\ref{ScalingBreaking}. Such a setting resembles
certain cases
discussed in ref. \cite{Amendola:2000uh}.

We expect that similar features
can also arise from a qualitative change of the scalar kinetic terms
for $\varphi>\varphi_0$. Let us recall, however, that the dark matter
- dark energy crossover may also be induced by a cosmological trigger
event \cite{Amendola:2007yx, Wetterich:2003kr}. In this case no particular feature for the scalar
potential or the kinetic terms at $\varphi = \varphi_0$ is needed, as for the leaping kinetic term of ref \cite{Hebecker:2000au}.

\section{Discussion}
In summary, we have found rather simple models for coupled cosmon
and bolon scalar fields which realize a consistent cosmology without
the need of other dark matter particles. The essential ingredients are
the presence of a period where the bolon energy density is much smaller
than radiation, and the increase of the bolon mass relative to the
Hubble parameter due to $\beta \ll \alpha$. 

Remarkably, the inverse
bolon mass in the present cosmological period,
\beq
\label{bolonmass}
m_\chi^{-1} = \sqrt{\frac{1}{3}} \frac{\chi_{\rm eq}}{M} \, H_{\rm eq}^{-1}
\, e^{
\beta \Delta \varphi / M} \approx \left( \frac{10 \, \chi_0}{M} \right)^4 \, {\rm
  pc} \, ,
\eeq
is typically found at subgalactic scales and could reduce the
clustering of dark matter on these and smaller scales \cite{Sahni:1999qe,
 Matos:2000ss}. Here $\Delta \varphi = \varphi_{\rm today} -
\varphi_{\rm eq}$ and the last factor is neglected in the weak-coupling approximation. We found that for the type of models discussed in section \ref{sec5} values of $\chi_0 / M \approx 1$ or somewhat larger are naturally realized, independently of the precise initial conditions. (A bolon mass of a galactic scale of $10$ kpc is realized for $\chi_0 / M \approx 1$.)

Since a coupling of the
bolon to ordinary matter has presumably at most gravitational strength,
a direct detection by the searches for WIMP-like particles or axions
seems excluded. On the other hand, a coupling $\beta$ between dark
energy and dark matter is a generic feature of our setting. It
reflects the common origin of the cosmon-bolon potential and the cosmon
and bolon masses from the deviations from dilatation symmetry. Interestingly, if cosmological measurements should indicate an equation of state w$_{\rm de} < -1$ for uncoupled dark energy, the
coupling $\beta$ can be used to explain such observations \cite{Das:2005yj}. Furthermore $\beta$ influences both the behaviour of the cosmological
solution and the properties of dark matter on smaller length scales. A test of the 
interesting observational consequences can constrain $\beta$ or give hints in the direction of the coupled dark energy and dark matter of our model.

\acknowledgments{SN is partially supported by the Academy of Finland
grant 136600.}

\end{document}